# Effect of growth oxygen pressure on anisotropic-strain-induced phase separation in epitaxial La$_{0.67}$Ca$_{0.33}$MnO$_3$/NdGaO$_3$(001) films

Bowen Zhi, Guanyin Gao, Zhen Huang, Lingfei Wang, Xuelian Tan, Pingfan Chen, and Wenbin Wu[a]

*Hefei National Laboratory for Physical Sciences at Microscale, and High Magnetic Field Laboratory, Chinese Academy of Sciences, University of Science and Technology of China, Hefei 230026, P.R. China*

**Abstract**

The effect of deposition oxygen pressure ($P_O$) on phase separation (PS) induced in epitaxial La$_{0.67}$Ca$_{0.33}$MnO$_3$/NdGaO$_3$(001) films was investigated. Fully oxygenated films grown at high $P_O$ are anisotropically strained. They exhibit PS over a wide temperature range, because of the large orthorhombicity of NdGaO$_3$ substrates. The paramagnetic insulator-to-ferromagnetic metal (FM) and FM-to-antiferromagnetic insulator (AFI) transitions gradually shift to lower temperatures with decreasing $P_O$. The AFI state is initially weakened ($P_O \geq 30$ Pa), but then becomes more robust against the magnetic field ($P_O < 30$ Pa). The out-of-plane film lattice parameter increases with decreasing $P_O$. For films grown at $P_O \geq 30$ Pa, the slight oxygen deficiency may enlarge the lattice unit cell, reduce the anisotropic strain and suppress the AFI state. Films deposited at $P_O < 30$ Pa instead experience an average compressive strain. The enhanced compressive strain and structural defects in the films may lead to the robust AFI state. These results aid our understanding of PS in manganite films.

[a] Electronic mail: wuwb@ustc.edu.cn.

**I. INTRODUCTION**

Epitaxial strain and the strong electron-lattice coupling can be used to tune the ground state of perovskite-manganite films. Resulting films exhibit properties including phase separation (PS) and low-field magnetoresistance, and have potential application in spintronic devices.[1-5] Ideal $ABO_3$ perovskites such as $SrTiO_3$ exhibit cubic symmetry with space group *Pm3m*. However, the symmetry of most perovskite-manganites is lower, e.g. orthorhombic or rhombohedral, because of tilt and rotation of $BO_6$ octahedra arising from the divergence of ionic radii between A and B, or alternatively because of Jahn-Teller (JT) distortion.[6] Increasing orthorhombicity leads to JT distortion, charge-orbital ordering (COO) and superexchange (SE) interaction, all of which strongly effect the manganite properties.[7,8] Using epitaxial strain to control the film structure can tune the electronic state of these correlated oxides.[6,9-13] It is straightforward and relatively efficient compared with doping or other reported methods. Manganites commensurately grown on different substrates (or on identical substrates but in differing orientation) will be subjected to lattice constant or symmetry mismatch. Structures will experience isotropic or anisotropic strain, and $BO_6$ octahedra will distort or tilt, affecting the films' electronic states.[10-14]

Changing the oxygen content by controlling the deposition oxygen pressure ($P_O$) can also tune the properties of manganite films. This is because of manganite's tendency to favor nonstoichiometric oxygen content.[15-17] Manganite films grown at lower $P_O$ are generally more oxygen-deficient. This affects the carrier density and also enlarges the lattice unit cell volume, thus changing the film's strain state. Oxygen deficiency in $La_{0.67}Sr_{0.33}MnO_3$ films has been reported to cause COO, because of the $MnO_6$ octahedral distortion.[15] The presence or absence of oxygen in Mn-O-Mn bonds strongly affects the double-exchange (DE) and SE interactions, giving rise to the coexistence and competition of different phases.[18,19]

Post-annealing under an oxidizing or reducing atmosphere also affects the oxygen content in manganite films.[20-27] Oxygen annealing can eliminate oxygen deficiency and optimize the film's oxygen content. This leads to enhanced DE interaction and paramagnetic insulating-to-ferromagnetic metal (FM) transition temperature $T_P$.[12] The structure and morphology of films can also be greatly improved by post-annealing in an oxygen atmosphere.[20-22] Oxygen deficiency can be imparted by vacuum annealing. Generated oxygen vacancies appear less stable than those in films deposited at low $P_O$, and can even be removed at room temperature under ambient atmosphere.[25] Oxygen diffusion has been shown to depend strongly on film strain states, which partly arises from the different substrates. A larger strain makes it more difficult for oxygen to diffuse in the film.[23,24] We recently reported on oxygen annealed $La_{0.67}Ca_{0.33}MnO_3/NdGaO_3(001)$ [LCMO/NGO(001)] films accompanied with the contraction of lattice unit cells. Anisotropically strained films were found to exhibit enhanced orthorhombicity and JT distortion, producing a tunable antiferromagnetic-insulator (AFI) phase over a wide temperature range at low magnetic field.[27]

In the current study, the effect of $P_O$ on LCMO/NGO(001) film PS is investigated. A large orthorhombic lattice distortion is exerted by the substrate. Films grown at high $P_O$ are fully oxygenated, anisotropically strained and exhibit PS over a wide temperature range. The paramagnetic-to-FM and FM-to-AFI transitions shift to lower temperatures with decreasing $P_O$. The AFI state is initially weakened ($P_O \geq 30$ Pa), but then becomes more robust against the magnetic field ($H$), when grown at $P_O < 30$ Pa. The oxygen deficiency controlled by $P_O$, and the film strain variation were analyzed. These observations aid our understanding of strain and PS in thin manganite films.

## II. EXPERIMENTS

LCMO films optimally doped for the FM ground state were ~40 nm thick. They were prepared on orthorhombic NGO(001) and NGO(110) substrates by pulsed laser deposition. The ceramic LCMO target was prepared by standard solid state reactions. During deposition, the laser energy density and repetition rate were constant at 2 J/cm$^2$ and 5 Hz, respectively. $P_O$ was 15-45 Pa, and the substrate temperature was 735 °C. After deposition, each film was annealed *in situ* for 15 min, before being cooled to room temperature. During this process, the O$_2$ pressure remained at $P_O$. To optimize the oxygen content, all films were annealed *ex situ* at 780 °C under an $O_2$ flow for 10 h.

Magnetoresistance and magnetization measurements were conducted on Quantum Design superconducting quantum interference device (SQUID) magnetometer (MPMS) and physical property measurement system (PPMS). Film structures were characterized by high-resolution X-ray diffraction (XRD), including the reciprocal space maps (RSMs), using CuK$\alpha_1$ ($\lambda$ = 1.5406 Å) radiation (PANalytical X'pert). Film thicknesses were determined by analyzing the Laue fringes around the high-angle diffraction peak. Film surface morphologies were analyzed by atomic force microscopy (AFM, Vecco, MultiMode V).

## III. RESULTS AND DISCUSSION

In *Pbnm* orthorhombic notation, the NGO substrate has lattice constants $a_S$ = 5.4332 Å, $b_S$ = 5.5034 Å, and $c_S$ =7.7155 Å, while bulk LCMO has those of $a_B$ = 5.4717 Å, $b_B$ = 5.4569 Å, and $c_B$ = 7.7112 Å. Thus, fully oxygenated LCMO/NGO(001) films experience a large anisotropic strain {−0.70% along [100] (compressive) and +0.85% along [010] (tensile)} from the substrate. The average in-plane lattice mismatch is negligible (+0.15%).[27] The oxygen deficiency in films deposited at much lower $P_O$ (although still post annealed in the same oxidizing atmosphere) cannot be ameliorated. The films' structural and physical properties differ greatly from those of fully oxygenated films.

Figure 1(a) shows XRD linear scans around the (002) reflections from films deposited at $P_O$ ranging from 45 to 15 Pa. For simplicity, scans are indexed in pseudocubic notation. Films deposited at 45 to 20 Pa are highly crystalline and smooth, as reflected by their clear Laue fringes. The LCMO(002) reflection gradually shifts to lower Bragg angle with decreasing $P_O$, corresponding to an increased out-of-plane lattice constant. Oxygen deficiency can enlarge the lattice unit cell. This observation implies that films remain oxygen-deficient after annealing, and that films deposited at lower $P_O$ contain more oxygen vacancies.[17,23] XRD RSMs on (103) reflections (also indexed in pseudocubic notation) from the film and substrate were measured to further understand this structure evolution. Figure 1(b) shows RSMs from films deposited at $P_O$ from 45 to 20 Pa. All films retain the same in-plane lattice spacing ($a = -\lambda/2Q_{[100]}$) as the substrate, indicating that they were coherently deposited on NGO(001) substrates. The out-of-plane lattice constant of the films ($c = 3\lambda/2Q_{[001]}$) increases with decreasing $P_O$ from 45 to 20 Pa, in agreement with the XRD scans in Fig. 1(a). The LCMO(103) reflection shifts from the higher to lower side of NGO(103), as $P_O$ decreases from 30 to 25 Pa. It can be deduced that the anisotropic strain is gradually reduced with decreasing $P_O$, and the average in-plane lattice strain may even change from tensile to compressive. The LCMO(002) reflection is broadened for the 15 Pa deposited film, implying that the oxygen distribution or strain field along the out-of-plane direction may be inhomogeneous. Figure 1(c) shows AFM images obtained from the same films, confirming that all annealed films have a smooth surface.

Figures 2(a) and 2(b) show temperature-dependent resistivity ($\rho$-$T$) curves, measured at zero-field from films deposited at various $P_O$. Films deposited at 45 and 35 Pa exhibit similar paramagnetic-to-FM transitions at the Curie temperature ($T_C \cong T_P$) 264 K. These transitions are the same as those of the bulk counterpart, indicating that these films are fully oxygenated. $T_P$ decreases to 246 K as $P_O$ is reduced to 30 Pa, which is attributed to a slight oxygen deficiency weakening the DE.[15] The film deposited at 45 Pa exhibits an "overshot" hysteresis with the AFI state onset at 255 K, and another metal-insulator transition (MIT) at 92 K (137 K) upon cooling (warming). This suggests that PS of the AFI and FM states has been induced in the films, which is in good agreement with our previous observations.[27] Films grown at 35 and 30 Pa exhibit a gradually depressed "overshot" hysteresis, suggesting a weakened AFI state. The AFI onset temperature is indicated by dashed lines in Figs. 2(c)-(e), and decreases slightly from 256 to 253 to 224 K, as $P_O$ reduces from 45 to 35 to 30 Pa, respectively. This implies that the AFI onset temperature is related to anisotropic strain, which in turn is controlled by the oxygen deficiency. The $\rho$-$T$ behavior for films deposited at $P_O < 30$ Pa is quite different. The $T_P$ of films grown at 25, 20 and 15 Pa decreases rapidly from 229 to 183 to 153 K, respectively. At the same time, the high-temperature AFI phase disappears, and another MIT at ~89 K appears. The onset temperature of this insulating state is less sensitive to $P_O$, but the residual resistivity increases dramatically as $P_O$ decreases. A lower $T_P$ followed by insulating behavior has been reported for other manganite films, and is generally attributed to biaxial epitaxial strain or oxygen-deficiency-induced COO at low temperature.[16,18]

It is clear from the zero-field $\rho$-$T$ curves that $P_O$ has a dramatic effect on $T_P$ and PS in LCMO/NGO(001) films. The magnetoresistance (MR) of these films during a field-cooling/field-warming (FC-FW) process is now investigated. The magnetic field $H$ is applied parallel to the film-plane. Figures 2(c)-(h) show that the $T_P$ increases with increasing $H$, resulting in a negative MR around $T_P$ (typical of doped manganites). For films deposited at 45, 35 and 30 Pa, the induced AFI state can be completely melted at $H > 3$, 1, and 0.5 T, respectively. That means the AFI state destabilizes as $P_O$ approaches 30 Pa. The AFI phase (with onset temperature near 89 K) for films grown at $P_O < 30$ Pa is much more robust against $H$. Figures 2(f)-(h) show that the AFI state of films deposited at 25, 20 and 15 Pa cannot be completely melted, at least at 7 T.

To further probe film phase instability, $\rho$-$T$ curves were measured during the field-cooling/zero-field-warming (FC-ZFW) process. Isothermal $\rho$-$H$ curves of the 45 and 15 Pa deposited films were also measured.[27,28] Figure 3(a) shows that after FC the 45 Pa film to 10 K at 7 T, the film retains the FM state till 108 K upon ZFW. This is followed by a sharp jump in $\rho$ of nearly five orders of magnitude, suggesting a recovered AFI state.[29,30] The FM state in this PS situation dominates at the lower temperature range, while the AFI state dominates at higher temperatures. Figure 3(b) shows $\rho$-$H$ curves from the 45 Pa deposited film. After zero-field-cooling (ZFC), the AFI state at 150 K fully melts at 3.52 T. The state then reforms as $H$ is decreased to 0.90 T during $H$ cycling. The AFI state is melted at 1.60 T (2.61 T) after ZFC the film to 50 K (10 K). A reformed AFI phase is not observed. The higher melting field at 10 K is indicative of a frozen PS state at low temperatures.[28] These observations are consistent with the dynamic $T$-$H$ phase diagram previously reported for fully oxygenated films.[27,29] Figure 3(c) shows that no steady FM state is observed upon ZFW when the film is deposited at 15 Pa, after FC to 10 K at 7 T. $\rho$ begins to increase upon removing $H$ at 10 K, and then merges with the ZFC-ZFW curve near 60 K. This implies the AFI state dominates the entire PS regime, and that the potential barrier between the FM and AFI states is relatively low.[30] Figure 3(d) shows $\rho$-$H$ curves of the 15 Pa deposited film. After ZFC the film to 150 K, MR behavior typical of a FM state is observed. At 10 K, most of the AFI state can be melted at 9 T, and the PS state appears difficult to freeze.[28] The AFI state in the 45 Pa deposited film melts sharply, while that in the 15 Pa deposited film does so sluggishly. This indicates the AFI state in highly oxygen-deficient films is more robust, and cannot melt homogeneously. The AFI states in these two samples are quite different in nature.

The as-grown 45 and 15 Pa deposited films were also examined, to clarify the post-annealing effect on forming different AFI states. In Fig. 4(a), the 45 Pa deposited as-grown film exhibits a FM ground state with $T_P$ at 240 K, indicating slight oxygen deficiency in the film. After post-annealing, the annihilated oxygen defects enhance the orthorhombicity of the commensurate film. This leads to both PS and an enhanced $T_P$. Figure 4(b) shows an increased LCMO(002) Bragg angle, with depressed Laue fringes at higher angles. This can indicate oxygen incorporation and accompanied lattice distortion. The lattice unit cells of as-grown films expand with decreasing $P_O$. For films grown at $P_O \geq 30$ Pa, this expansion is moderate, and the enlarged lattice unit cell reduces the average lattice mismatch. This leads to decreased anisotropic strain and even the disappearance of the high-temperature AFI state. For films grown at $P_O < 30$ Pa, this expansion is rather large. The average lattice strain changes from tensile to compressive, as seen in Figs. 4(b) and 4(d). During post-annealing, the larger compressive strain could hinder oxygen in-diffusion, as diffusion depends strongly on the strain state.[23,24] Figure 4(c) shows the $\rho$-$T$ curve of the 15 Pa deposited film. The as-grown film has a $T_P$ of 129 K, which is much lower than that of its bulk counterpart, indicating a large amount of oxygen vacancies in the film. $T_P$ increases to only 153 K after annealing. The AFI state sharpens at low temperature, due to a large biaxial strain or the presence of oxygen-deficiency-induced COO. Figure 4(d) shows the broadened LCMO(002) reflection and weakened Laue fringes, which suggests the annealed film is less uniform than the as-grown one. The enhanced compressive strain and structural defects (oxygen deficiency, inhomogeneous strain field) should be considered for the formation of this AFI state.[18,19]

Another possible cause for the AFI state emerging in films (doped nominally for a FM ground state) is the variation in La:Ca doping ratio or the cation distribution. These factors lead to a complicated phase diagram or phase coexistence.[4,5] During laser deposition at low $P_O$, the cation ratio may deviate and change the films' physical properties. To probe this possibility, $\rho$-$T$ curves and XRD linear scans around the (220) reflection from the (110)-oriented LCMO/NGO(110) films were measured, as shown in Fig. 5. Figure 5(a) shows that $T_P$ decreases from 265 to 247 K, when $P_O$ decreases from 45 to 15 Pa. All films exhibit a FM ground state, quite different from the situation with LCMO/NGO(001) films. All films are ~40 nm thick and were deposited and annealed side by side in the same run, so the La:Ca ratio should be constant. This inference is further confirmed by the LCMO(220) reflection gradually shifting to lower Bragg angles, as $P_O$ is decreased from 45 to 15 Pa. This is shown in Fig. 5(b), and is similar to the behavior of LCMO/NGO(001) films. The absence of an AFI state in fully oxygenated (110)-films has been previously attributed to negligible lattice strain and fixed bulk-like orthorhombicity in this special orientation (i.e. no clamping of the $a$, $b$ axes in orthorhombic notation).[27] Thus, the variation in La:Ca ratio, if any, is not expected to play a dominant role in emerging AFI states in LCMO/NGO(001) films.

## IV. SUMMARY

The effect of $P_O$ on PS induced in epitaxial LCMO/NGO(001) films has been investigated. Films grown at high $P_O$ are anisotropically strained, and exhibit PS over a wide temperature range, because of the large substrate orthorhombicity. $T_P$ and the film FM-to-AFI transition shift to lower temperature with decreasing $P_O$. The AFI state is initially weakened ($P_O \geq 30$ Pa), but then becomes more robust against $H$ when grown at $P_O$ < 30 Pa. At $P_O$ ≥ 30 Pa, the slight oxygen deficiency can enlarge the lattice unit cell, reduce the anisotropic strain and suppress the AFI state in the films. At $P_O$ < 30 Pa, however, a large compressive strain arises. The high oxygen deficiency, changed strain field and inhomogeneous film distribution may lead to the robust AFI state at low temperature.

**ACKNOWLEDGEMENTS**

This work was supported by the NSF of China (Grant Nos. 11074237, 11274287) and the National Basic Research Program of China (Grant Nos. 2009CB929502, 2012CB927402).

1. S. Jin, T. H. Tiefel, M. Mc Cormack, R. A. Fastnacht, R. Ramesh, and L. H. Chen, Science **264**, 413 (1994).

2. A. J. Millis, Nature **392**, 147 (1998).

3. A. Sadoc, B. Mercey, C. Simon, D. Grebille, W. Prellier, and M. –B. Lepetit, Phys. Rev. Lett. **104**, 046804 (2010).

4. K. Dorr, J. Phys. D **39**, R125 (2005).

5. E. Dagotto, T. Hotta, and A. Moreo, Phys. Rep. **344**, 1 (2001).

6. J. M. Rondinelli, S. J. May, and J. W. Freeland, MRS Bulletin **37**, 261 (2012).

7. H. Kajimoto, H. Yoshizawa, H. Kawano, H. Kuwahara, Y. Tokura, K. Ohovama, and M. Ohashi, Phys. Rev. B **60**, 9506 (1999).

8. A. M. Haghiri-Gosnet, M. Hervieu, Ch. Simon, B. Mercey, and B. Raveau, J. Appl. Phys. **88**, 3545 (2000).

9. K. H. Ahn, T. Lookman, and A. R. Bishop, Science **428**, 401 (2004).

10. K. J. Lai, M. Nakamura, W. Kundhikanjana, M. Kawasaki, Y. Tokura, M. A. Kelly, Z. X. Shen, Science **329**, 190 (2010).

11. Amlan Biswas, M. Rajeswari, R. C. Srivastava, T. Venkatesan, , R. L. Greene, Q. Lu, A. L. de Lozanne and A. J. Millis, Phys. Rev. B **63**, 184424 (2001).

12. W. Prellier, M. Rajeswari, T. Venkatesan, and R. L. Greene, Appl. Phys. Lett. **75**, 1446 (1999).

13. Y. Z. Chen, J. R. Sun, S. Liang, W. M. Lv, B. G. Shen, and W. B. Wu, J. Appl. Phys. **103**, 096105 (2008).

14. Dane Gillaspie, J. X. Ma, Hong-Ying Zhai, T. Z. Ward, Hans M. Christen, E. W. Plummer, and J. Shen, J. Appl. Phys. **99**, 08S901 (2006).

15. J. Li, C. K. Ong, J. M. Liu, Q. Huang, and S. J. Wang, Appl. Phys. Lett. **76**, 1051 (2000).

16. J. Sakai, N. Ito, and S. Imai, J. Appl. Phys. **99**, 08Q318 (2006).

17. Anjali S. Ogale, S. R. Shinde, V. N. Kulkarni, J. Higgins, R. J. Choudhary, Darshan C. Kundaliya, T.

Polleto, S. B. Ogale, R. L. Greene, and T. Venkatesan, Phys. Rev. B **69**, 235101 (2004).

18. H. Ni, D. Yu, K. Zhao, Y.-C. Kong, S. Q. Zhao, and W. S. Zhang, J. Appl. Phys. **110**, 033112 (2011).

19. T. Petrisor, Jr., M. S. Gabor, A. Boulle, C. Bellouard, C. Tiusan, O. Pana, and T. Petrisor, J. Appl. Phys. **109**, 123913 (2011).

20. S. H. Seo, H. C. Kang, H. W. Jang, and D. Y. Noh, Phys. Rev. B **71**, 012412 (2005).

21. P. Murugavel, J. H. Lee, K. –B. Lee, J. H. Park, J. –S. Chung, J. –G. Yoon, and T. W. Noh, J. Phys. D **35**, 3166 (2002).

22. V. G. Prokhorov, V. A. Komashko, and V. L. Svetchnikov, Phys. Rev. B **69**, 014403 (2004).

23. J. R. Sun, C. F. Yeung, K. Zhao, L. Z. Zhou, C. H. Leung, H. K. Wong, and B. G. Shen, Appl. Phys. Lett. **76**, 1164 (2000).

24. B. Li, L. Yang, J. Z. Tian, X. P. Wang, H. Zhu, and T. Endo, J. Appl. Phys. **109**, 073922 (2011).

25. Z. Z. Yin, G. Y. Gao, Z. Huang, X. L. Jiao, Q. Z. Liu, and W. B. Wu, J. Phys. D: Appl. Phys. **42**, 125002 (2009).

26. M. Egilmez, K. H. Chow, and J. Jung, Appl. Phys. Lett. **89**, 062511 (2006).

27. Z. Huang, G. Y. Gao, Z. Z. Yin, X. X. Feng, Y. Z. Chen, X. R. Zhao, J. R. Sun, and W. B. Wu, J. Appl. Phys. **105**, 113919 (2009); Z. Huang, L. F. Wang, P. F. Chen, G. Y. Gao, X. L. Tan, B. W. Zhi, X. F. Xuan and W. B. Wu, Phys. Rev. B **86**, 014410 (2012).

28. W. Wu, C. Israel, N. Hur, S. Park, S.-W. Cheong, and A. D. Lozanne, Nature Mater. **5**, 881 (2006).

29. Z. Huang, L. F. Wang, X. L. Tan, P. F. Chen, G. Y. Gao, and W. B. Wu, J. Appl. Phys. **108**, 083912 (2010).

30. H. Kuwahara, Y. Moritomo, Y. Tomioka, A. Asamitsu, M. Kasai, R. Kumai, and Y. Tokura, Phys. Rev. B **56**, 9386 (1997).

FIG. 1. (Color online) (a) XRD linear scans around NGO(002) (sharp peak at $2\theta = 47.074°$) and LCMO(002) reflections from LCMO/NGO(001) films deposited at different $P_O$. Scans are shown offset for clarity. (b) XRD RSMs on (103) reflections from films deposited at 45-20 Pa. Arrows indicate the central position of NGO(103) and LCMO(103) along the out-of-plane direction. (c) AFM images from films deposited at 45-20 Pa.

FIG. 2. (Color online) $\rho$-$T$ curves measured at zero-field from LCMO/NGO(001) films deposited at (a) 45, 35 and 30 Pa and (b) 25, 20 and 15 Pa. In (c)-(h), films were measured under various magnetic fields.

FIG. 3. (Color online) FC-ZFW $\rho$-$T$ curves [(a) and (c)] and isothermal $\rho$-$H$ curves [(b) and (d)] measured from films deposited at 45 and 15 Pa, respectively. FC-ZFW $\rho$-$T$ curves were measured after FC the film from 320 to 10 K at 7 T. $\rho$-$H$ curves were measured after ZFC the film to the stated temperature. ZFC-ZFW $\rho$-$T$ curves are also shown in (a) and (c) for comparison.

FIG. 4. (Color online) ZFC-ZFW $\rho$-$T$ curves from as-grown and annealed films deposited at (a) 45 and (c) 15 Pa. XRD linear scans around the LCMO (002) reflections of these films are shown in (b) and (d), respectively.

FIG. 5. (Color online) (a) ZFC-ZFW $\rho$-$T$ curves from annealed LCMO/NGO(110) films deposited at 45, 30, 20 and 15 Pa. (b) XRD linear scans around the NGO(220) and LCMO(220) reflections from LCMO/NGO(110) films deposited at different $P_O$. Scans are shown offset for clarity.

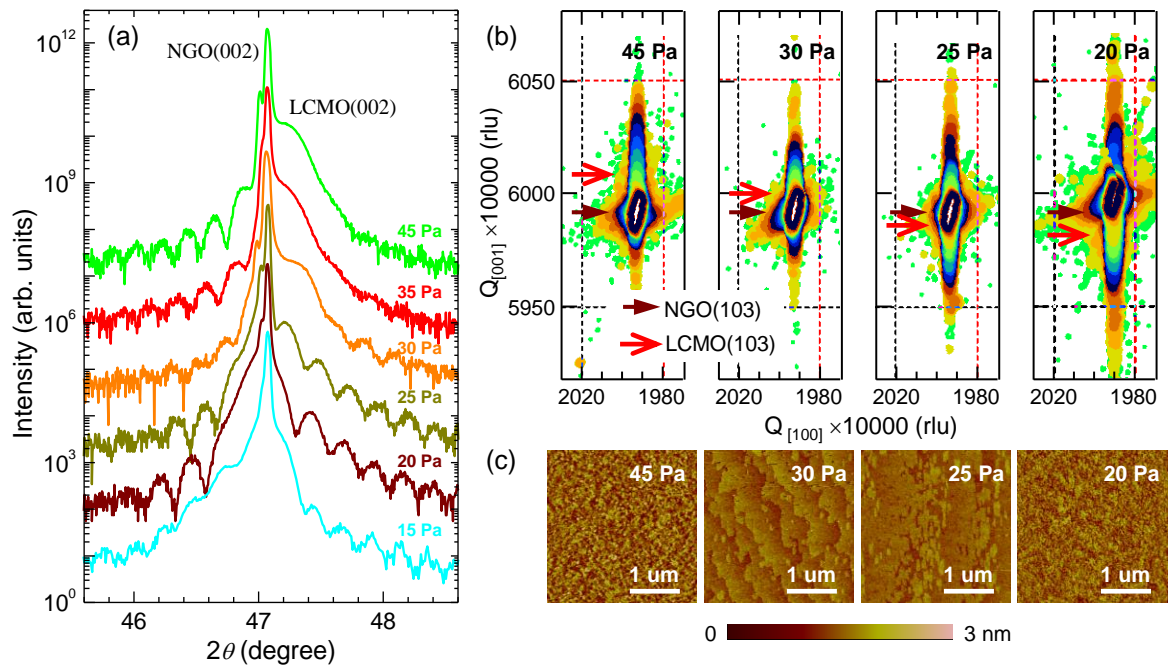

Fig. 1

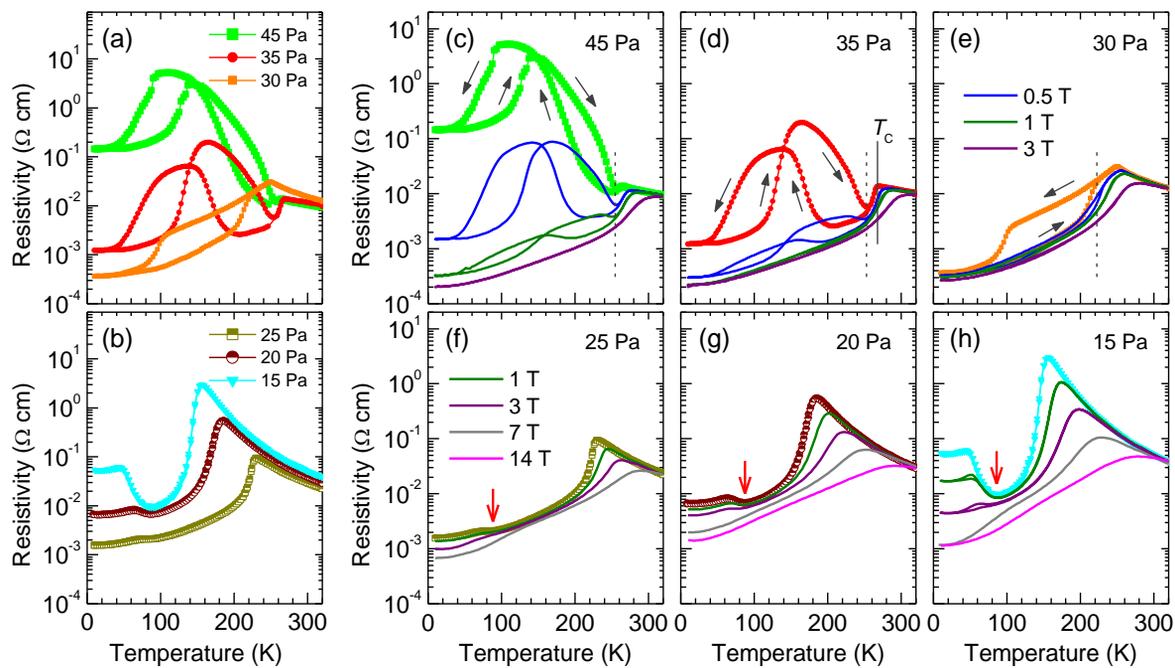

Fig. 2

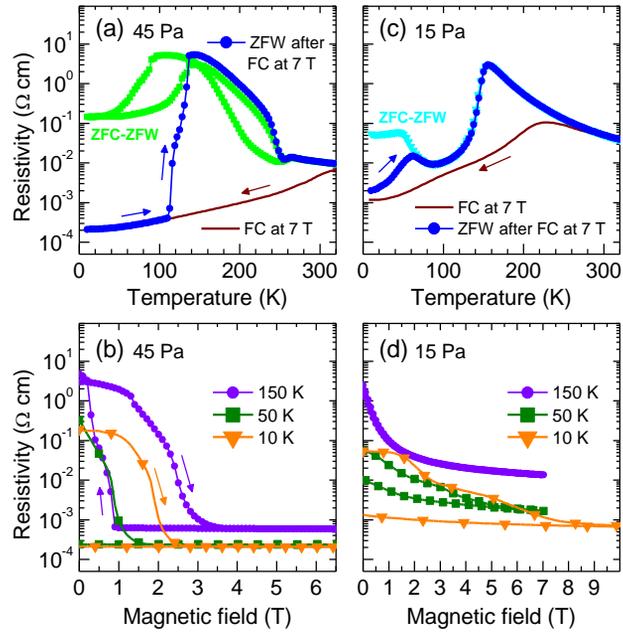

Fig. 3

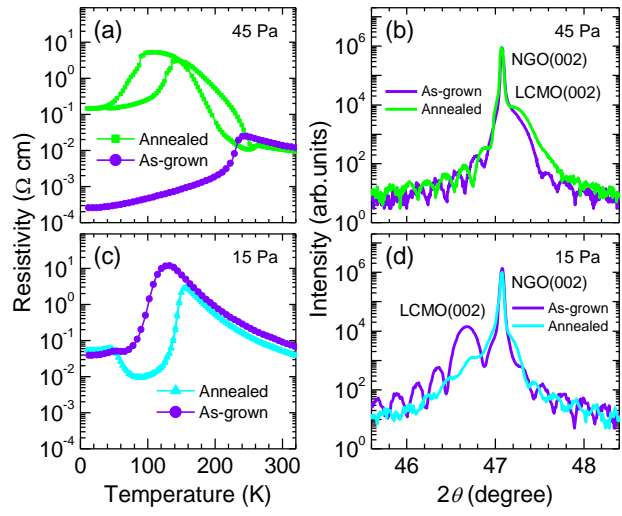

Fig. 4

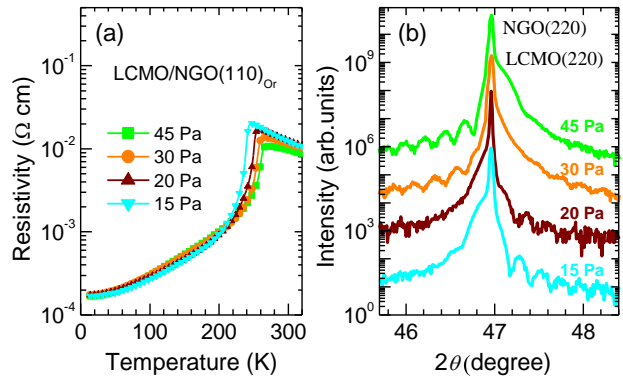

Fig. 5